# Heat Transfer in Buildings: Application to Solar Air Collector and Trombe Wall Design

H. Boyer, F. Miranville, D. Bigot, S. Guichard, I. Ingar,
A. P. Jean, A. H. Fakra, D. Calogine and T. Soubdhan
*University of La Reunion,*
*Physics and Mathematical Engineering for Energy and Environment Laboratory,*
*LARGE - GéoSciences and Energy Lab., University of Antilles et de la Guyanne,*
*France*

## 1. Introduction

The aim of this paper is to briefly recall heat transfer modes and explain their integration within a software dedicated to building simulation (*CODYRUN*). Detailed elements of the validation of this software are presented and two applications are finally discussed. One concerns the modeling of a flat plate air collector and the second focuses on the modeling of Trombe solar walls. In each case, detailed modeling of heat transfer allows precise understanding of thermal and energetic behavior of the studied structures.

Recent decades have seen a proliferation of tools for building thermal simulation. These applications cover a wide spectrum from very simplified steady state models to dynamic simulation ones, including computational fluid dynamics modules (Clarke, 2001). These tools are widely available in design offices and engineering firms. They are often used for the design of HVAC systems and still subject to detailed research, particularly with respect to the integration of new fields (specific insulation materials, lighting, pollutants transport, etc.).

## 2. General overview of heat transfer and airflow modeling in *CODYRUN* software

### 2.1 Thermal modeling

This part is detailed in reference (Boyer, 1996). With the conventional assumptions of isothermal air volume zones, unidirectional heat conduction and linearized exchange coefficients, nodal analysis integrating the different heat transfer modes (conduction, convection and radiation) achieve to establish a model for each constitutive thermal zone of the building (one zone being a room of a group of rooms with same thermal behaviour).

For heat conduction in walls, it results from electrical analogy that the nodal method leads to the setting up of an electrical network as shown in Fig. 1, the number of nodes depending on the number of layers and spatial discretization scheme :

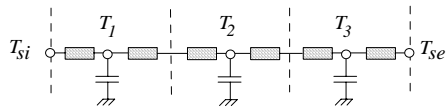

Fig. 1. Example of associated electrical network associated to wall conduction

The physical model of a room (or group of rooms) is then obtained by combining the thermal models of each of the walls, windows, air volume, which constitute what CODYRUN calls a zone. To fix ideas, the equations are of the type encountered below:

$$C_{si} \frac{dT_{si}}{dt} = h_{ci}(T_{ai} - T_{si}) + h_{ri}(T_{rm} - T_{si}) + K(T_{se} - T_{si}) + \varphi_{swi} \quad (1)$$

$$C_{se} \frac{dT_{se}}{dt} = h_{ce}(T_{ae} - T_{se}) + h_{re}(T_{sky} - T_{se}) + K(T_{si} - T_{se}) + \varphi_{swe} \quad (2)$$

$$C_{ai} \frac{dT_{ai}}{dt} = \sum_{j=1}^{Nw} h_{ci} S_j (T_{ai} - T_{si}(j)) + c\,\dot{m}(T_{ae} - T_{ai}) \quad (3)$$

$$0 = \sum_{j=1}^{Nw} h_{ri} A_j (T_{si}(j) - T_{rm}) \quad (4)$$

Equations of type (1) and (2) correspond to energy balance of nodes inside and outside surfaces. Nw denote the number of walls of the room, and correspond to the following figure:

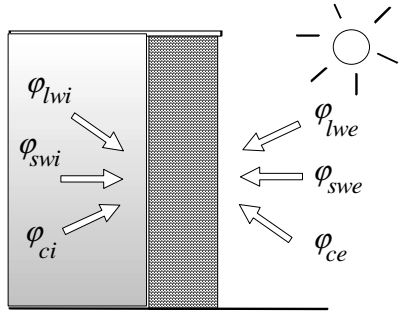

Fig. 2. Wall boundary conditions

$\varphi$ being heat flux density, *lw*, *sw*, and *c* indices refers to long wave radiation, short wave radiation and convective exchanges. Indices *e* and *i* are for exterior and indoor. Equation *(3)* comes from the thermo-convective balance of the dry-bulb inside air temperature $T_{ai}$, taking into account an air flow $\dot{m}$ through outside ($T_{ae}$) to inside. *(4)* represents the radiative equilibrium for the averaged radiative node temperature, $T_{rm}$. The generic approach application implies a building grid-construction focus. Following the selective model application logic developed (Boyer, 1999), an iterative coupling process is implemented to manage multi-zone projects. From the first software edition, some new models were implemented. They are in relation with the radiosity method, radiative zone coupling (short wavelength, though window glass) or the diffuse-light reduction by close solar masks (Lauret, 2001). Another example of *CODYRUN* evolution capacity could be illustrated by a specific study using nodal reduction algorithms (Berthomieu, 2003). To resolve it, a new implementation was done: the system was modified into a canonic form (in state space).

From a user utilization point of view, this thermal module exit is mainly linked to the discretizated elements temperatures, the surfacic heat flows, zone specific comfort indices, energy needs, etc.

## 2.2 Airflow modeling

Thermal and airflow modules are coupled to each other thanks to an iterative process, the coupling variables being the air mass flows. It is useful to implement the inter zones flows with the help of a matrix $\dot{m}$. The terms $\dot{m}[i, j]$ then qualifies the mass transfer from the zone $i$ to the zone $j$, as it is depicted .

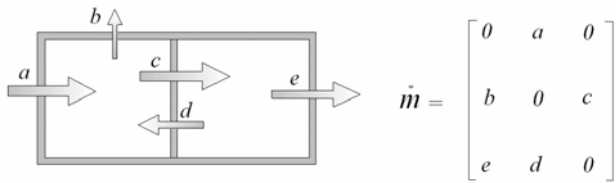

Fig. 3. Interzonal airflow rates matrix

Then, a pressure model for calculating previous airflow rates is set and takes into account wind effect and thermal buoyancy. To date, the components are integrated ventilation vents, small openings governed by the equation of crack flow, as :

$$\dot{m} = Cd\, S\, (\Delta P)^n \qquad (5)$$

$S$ is aperture surface, $Cd$ and $n$ (typically 0.5- 0.67) depends on airflow regime and type of openings and $\Delta P$ the pressure drop. Therefore, the following scheme can be applied for encountered cases.

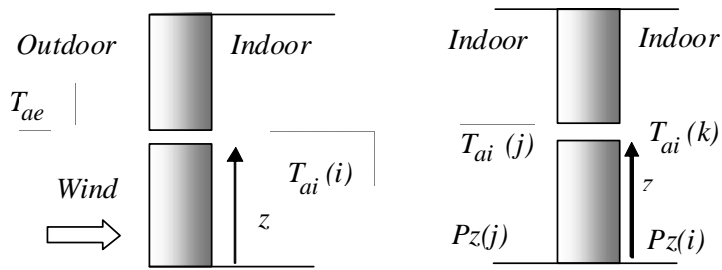

Fig. 4. Exterior and interior small openings

Taking into account the inside air volume incompressibility hypothesis, the mass weight balance should be nil. Thus, in the established system, called pressure system, the unknowns are the reference pressure for each of the zones. Mechanical ventilation is simply taken into account through its known airflow values. In this way, ventilation inlets are not taken into detailed consideration which could have been the case owing to their flow-pressure characteristics.

$$\begin{cases} \sum_{i=0, i \neq 1}^{i=N} \dot{m}(i,1) + \dot{m}_{vmc}(1) = 0 \\ \sum_{i=0, i \neq 2}^{i=N} \dot{m}(i,2) + \dot{m}_{vmc}(2) = 0 \\ \quad \quad \quad \ldots \\ \sum_{i=0}^{i=N-1} \dot{m}(i,N) + \dot{m}_{vmc}(N) = 0 \end{cases} \quad (6)$$

$\dot{m}_{vmc}(k)$ is the flow extracted from zone k (by vents) and N the total number of zones. After the setting up of this non linear system, its resolution is performed using a variant of the Newton-Raphson (under relaxed method). This airflow model has recently been supplemented by a $CO_2$ propagation model in buildings (Calogine, 2010).

### 2.3 Humidity transfers
For decoupling reasons, zones dry temperatures, as well as inter zone airflows, have been previously calculated. At this calculation step, for a given building, each zone's specific humidity evolution depends on the matricial equation :

$$C_h.r = A_h.r + B_h \quad (7)$$

According to Duforestel's model, an improvement of the hydric model (Lucas, 2003) was given by the use of a hygroscopic buffer. If we consider *N* as the number of zones, a linear system of dimension *2N* is obtained and a finite difference scheme is used for numerical solution.

### 3. Some elements of *CODYRUN* validation
#### 3.1 Inter model comparison
A large number of comparisons on specific cases were conducted with various softwares, whether CODYBA, TRNSYS for thermal part, with AIRNET, BUS, for airflow calculation or even CONTAM, ECOTECT and ESP. To verify the correct numerical implementation of models, we compared the simulation results of our software with those from other computer codes (Soubdhan, 1999). The benchmark for inter-comparison software is the BESTEST procedure (Judkoff, 1983), (Judkoff, 1995). This procedure has been developed within the framework of Annex 21 Task 12 of the Solar Heating and Cooling Program (SHC) by the International Energy Agency (IEA). This tests the software simulation of the thermal behavior of building by simulating various buildings whose complexity grows gradually and comparing the results with other simulation codes (ESP DOE2, SERI-RES, CLIM2000, TRNSYS, ...). The group's collective experience has shown that when a program showed major differences with the so-called reference programs, the underlying reason was a bug or a faulty algorithm.

### 3.1.1 IEA BESTEST cases

The procedure consists of a series of buildings carefully modeled, ranging progressively from stripped, in the worst case, to the more realistic. The results of numerical simulations, such as energy consumed over the year, annual minimum and maximum temperatures, peak power demand and some hourly data obtained from the referral programs, are a tolerance interval in which the software test should be. This procedure was developed using a number of numerical simulation programs for thermal building modeling, considered as the state of the art in this field in the US and Europe. After correction of certain parts of CODYRUN, an example output from this confrontation is the following concerns and energy needs for heating over a year, for which the results of our application are clearly appropriate.

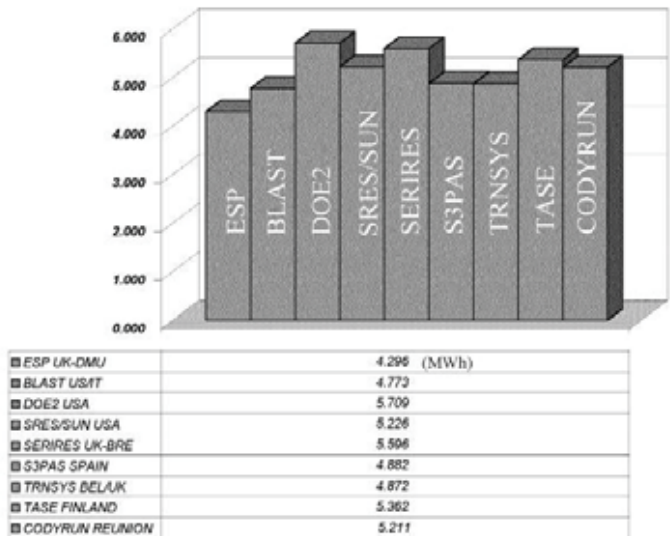

Fig. 5. Exemple of BESTEST result

The procedure requires more than a hundred simulations and cases treated with CODYRUN showed results compatible with most referral programs, except for a few. We were thus allowed to reveal certain errors, such as the algorithm for calculating the transmittance of diffuse radiation associated with double glazing and another in the distribution of the incident radiation by the windows inside the envelope. As in the reference codes, two points are to remember: the majority of errors came from a poor implementation of the code and all the efficiency of the procedure BESTEST is rechecked, errors belonging to modules that have been used for years.

Concerning airflow validation, IEA multizone airflow test cases describe several theoretical and comparative test cases for the airflow including multizone cases. It was developed within the International Energy Agency (IEA) programs: Solar Heating and Cooling (SHC) Task 34 and Energy Conservation in Building and Community Systems (ECBCS) Annex 43.

The tests are designed for testing the capability of building energy simulation programs to predict the ventilation characteristics including wind effect, buoyancy and mechanical fan. CODYRUN passed all the test cases.

### 3.1.2 TN51 airflow case

In (Orme, 1999), a test case composed of a building with three storeys is described and first published in (Tuomala, 1995). All boundary conditions are imposed (wind, external temperature) and indoor conditions are constant, i.e. in steady state.

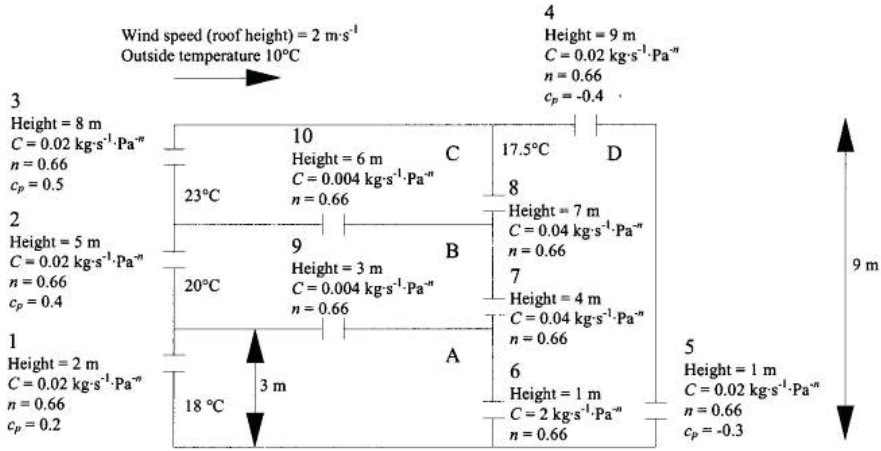

Fig. 6. TN 51 Test case

Next figure shows solution of 3 references codes, i.e. COMIS, CONTAM93 and BREEZE. CODYRUN's results were added on the figure extracted from the paper (Orme, 1999).

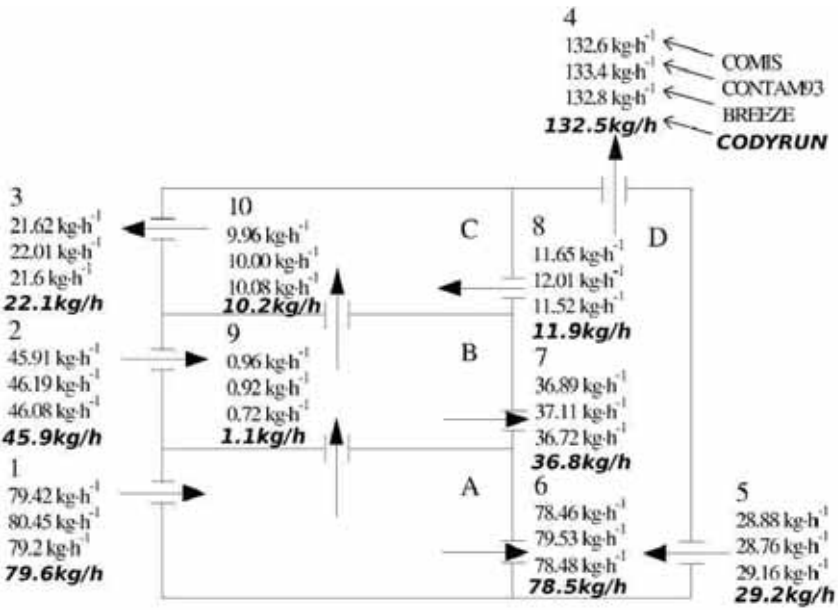

Fig. 7. Intermodel airflow comparison

As it can be seen, in terms of numerical results, *CODYRUN* gives nearly the same values as the other codes, little differences being linked to numerical aspects as algorithms or convergence criteria used.

**3.2 Experimental confrontation**
As part of measurement campaigns on dedicated cells and real homes, a set of elements of models have been to face action (and for some improved), mostly concerning thermal and humidity aspects. Not limited to, some of these aspects are presented in the articles (Lauret, 2001) (Lucas, 2001) and (Mara, 2001), (Mara, 2002).

**4. Two applications of the software**

**4.1 Air solar collector**
An air solar collector is used to heat air from solar irradiation. More precisely, it allows to convert solar energy (electromagnetic form) into heat (Brownian motion). This type of system can be used to heat buildings or to preheat air for drying systems. Air solar collector is constituted by an absorber encapsulated into an isolated set-up (Fig. 8). To avoid heat loss by the incoming flux side, a glass is located above the absorber, letting a gap between them. This glass allowing to catch long wavelength emitted by the set-up when its temperature rises-up. Air temperature improvement is mostly done by air convection at the absorber interface. This heat is then carried by air in the building or in the drying system.

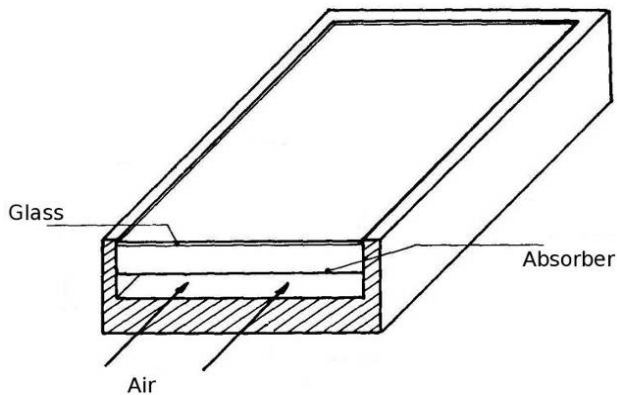

Fig. 8. Air solar collector principle, adapted from (Sfeir, 1984)

The enclosure is not strictly speaking a building, but the generic nature of the software allows to study this device. This is a zone (within the captation area), a glazed windows, exterior walls and an inner wall of absorber on which it is possible to indicate that solar radiation is incident.
It should be noted that this is part of a class simulation exercise allowing students to learn collector physics basis.

**4.1.1 Collector modeling and description of the study**

It is assumed that before initial moment, there is no solar irradiation and the collector is at thermal equilibrium, i.e. outside air temperature is the set-up temperature. At $t=0$, an irradiation flux Dh is applied without interactions with the outside air temperature. In a simplified approach, a theoretical study is easily obtained. Some equations below are used to understand the functioning of the collector. In steady state, the incoming irradiation flux reaching the absorber is fully converted and transmitted to the air (assumed to uniform temperature):

$$\rho_{as} C_{as} V \frac{dT_{ai}}{dt} = \tau_0 D_h - K_p S (T_{ai} - T_{ae}) - \dot{m} C_{as} (T_{ai} - T_{ae}) \quad (8)$$

Where $K_p$ is the thermal conductance (W.m$^{-2}$.K$^{-1}$) of each wall through which losses can pass, $m$ dot is a constant air mass flow through the collector, $T_{ai}$ is the inside air temperature, $T_{ae}$ is the outside air temperature, and $\rho_{as}$ (1.2 kg.m$^{-3}$) and $C_{as}$ (1000 J.kg$^{-1}$.K$^{-1}$) are respectively the air thermal conductivity and density. By knowing initial conditions (at $t=0$, $T_{ai}=T_{ae}$), the solution of equation (8) appears:

$$T_{ai}(t) = T_{ae} + \frac{\tau_0 D_h}{K} \left(1 - e^{-tK/\rho_{as} C_{as} V}\right) \quad (9)$$

The power delivered by the collector is given by

$$P_u = \dot{m} C_{as} (T_{ai} - T_{ae}) \quad (10)$$

The studied collector is horizontal, the irradiated face has a surface of 1m², and an air gap thickness of 0.1m. The meteorological file is constituted of a constant outside air temperature (25°C) and a constant direct horizontal solar irradiation of 500W (between 6 am and 7 pm). It is assumed (and further verified) that the steady state is reached at the end of the day.

Concerning the building description file, the collector is composed by 1 zone, 3 boundaries (1 for the upper face, 1 for the lower face, and 1 for all lateral faces), and 4 components (glass, lateral walls, lower wall and absorber). Walls are made - from inside to outside - of wood (1cm), polystyrene (5cm) and wood (1cm). Wood walls have a short wave absorptivity of α = 0.3.

The glass type is a single layer ($\tau_0$ = 0.85), and the absorber is a metallic black sheet of 2mm (α = 0.95).

The solar irradiation absorption model of the collector is a specific one. This has to be chosen in CODYRUN, because traditionally in buildings the solar irradiation doesn't impact directly an internal wall (absorber), but the floor.

Even if this solar air heater is considered as a very simple one, the following figure gives some details about heat exchanges in the captation area.

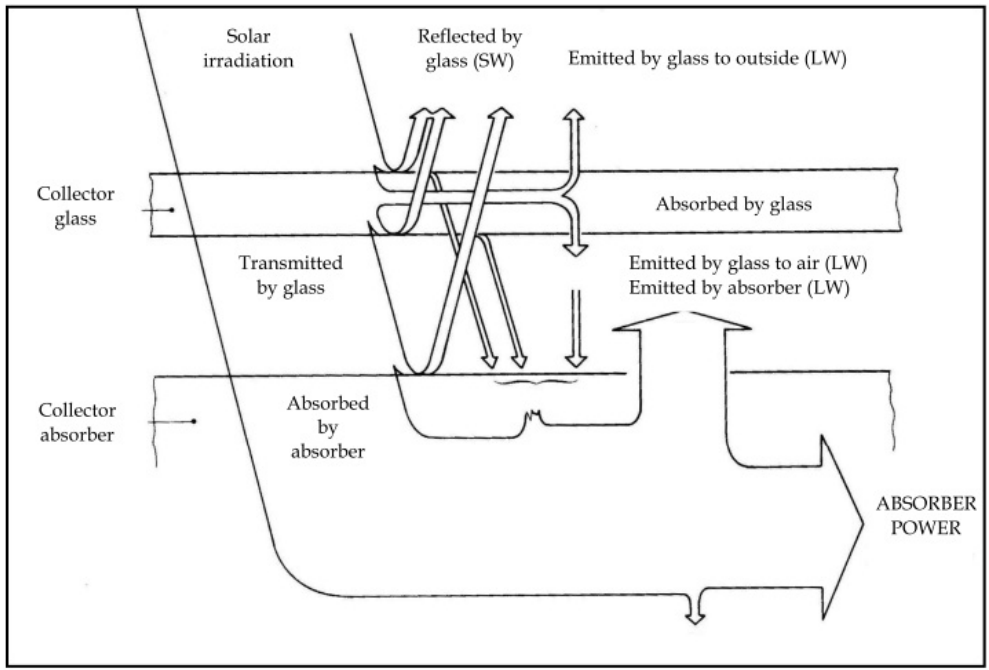

Fig. 9. Solar irradiation absorption process.

**4.1.2 Simulation results in the case of forced convection in air gaps**
An air flow of 1 m³.h⁻¹ is set. After running simulations, the following graph is obtained for two identical days :

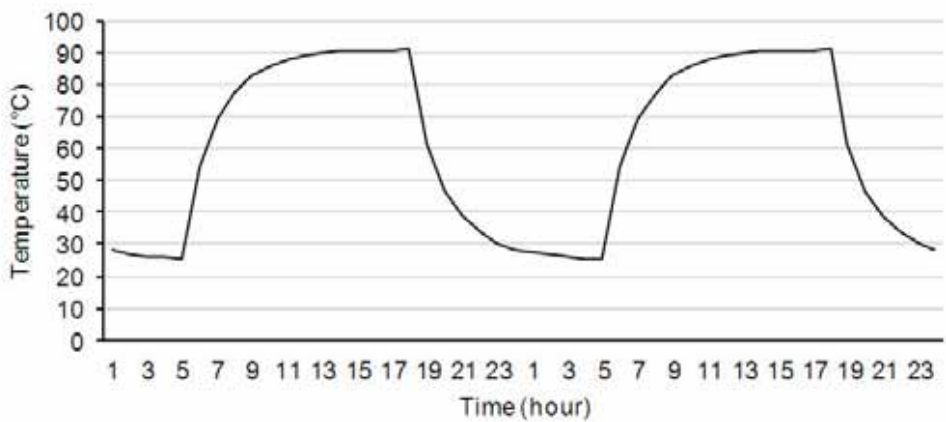

Fig. 10. Evolution of air temperature at the output of the solar collector with forced convection (1m³.h⁻¹).

Simulation results are consistent with simplified study, but not exactly identical. In fact, CODYRUN model is more precise than those use in eq. 8: thermal inertia of walls are taken into account, superficial exchanges are detailed, exchange coefficients are depending on inclination, glass transmittance is depending on solar incidence, …

With a low air flow rate of 1 m³.h⁻¹, the collector output power is about 22W and its efficiency about 4.5%. A solution to improve efficiency is to increase air flow rate (see Fig. 11), in link with the fact that conductive thermal losses are decreasing when internal temperature is decreasing.

Increase of flow rate leads to a lower air output temperature (and also to change convection exchange coefficients). In some cases like some drying systems where a minimal air output temperature should be needed, this can be problematic. In these cases, *CODYRUN* can help to choose the best compromise between air output temperature and collector power or efficiency.

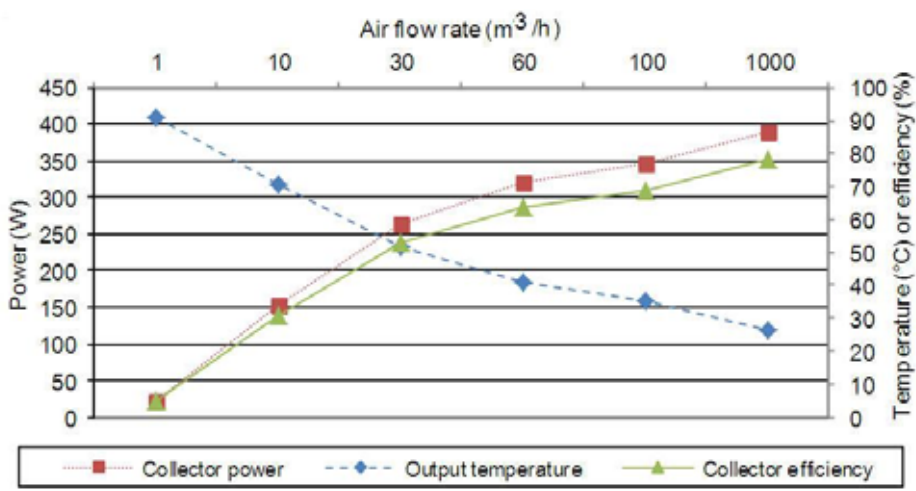

Fig. 11. Evolution of collector outputs regarding air flow rate.

Many other simulations could be conducted, in particular in the case of natural convection. Some of these cases could be applied to air pre-heating in case of passive building design, as for next application.

### 4.2 Trombe wall modeling
### 4.2.1 Trombe-Michel wall presentation
A Trombe-Michel wall (often called Trombe wall), is a passive heating system invented and patented (in 1881) by Edward Morse. It takes its name from two French people who popularized it in 1964, one engineer and one architect, respectively Félix Trombe and Jacques Michel.

A Trombe-Michel wall allows to store solar energy and distribute heat following simple physical principles. This system is composed of a vertical wall, submitted to the sunshine, which absorb and store the radiative energy as thermal energy by inertia. To improve its absorptivity, the wall can be paint in a dark color. Stored energy rises-up the wall temperature, inducing some heat transfer such as conduction (into the wall from outdoor to indoor side), radiation and convection (both with in/outdoor).

To avoid outdoor losses, a glazing, which is not transmissive to ultraviolet and far infrared, is located few centimeters in front of the wall (outdoor side). So the glazing reduces outside losses by convection and traps wall's radiation due to its temperature (far infrared emission) into the building. But it also slightly reduces the solar incoming flux, because some of it is reflected or absorbed.

*Nota*: Considering the glazing, the sunshine insulation (etc.), it is not easy to choose the optimal set-up configuration; this kind of situations are typical cases where *CODYRUN* simulations allow to find out.

Heat transfer through the wall is done by conduction. According to the wall size and its composition, a thermal delay appears between heat absorption and emission. This characteristic allows to configure specifically the system in function of the needs.

Considering the thermal delay, conductive transfer-mode is convenient to heat by night, but not by day. To palliate this lack, a faster transfer mode is brought by a variant of the Trombe-Michel wall, called recycling wall (Fig. 12). The recycling wall is obtained by holes addition at the top and the bottom of the wall. This variant allows to get a natural air flow, transferring energy by sensitive enthalpy between the system side of the wall (also called Trombe side) and the room. Indeed, as long as the glass is far enough of the wall, the temperature difference between them induces some air-convective movement, initiating the global air-circulation, so allowing heat distribution.

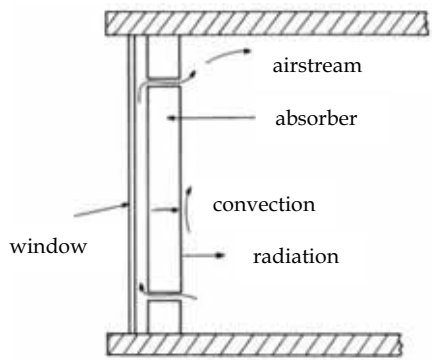

Fig. 12. Trombe-Michell recycling wall set-up (Sfeir, 1981)

Several variants of the Trombe-Michel wall can be used. These one usually evolve in function of the climatic need and can, for example, be defined by: the holes size and presence (or not), the characteristics of the storage wall (e.g. thickness, density, presence of fluid circulation or phase change materials, etc. ) or the glazing type (e.g. simple or double, treated surface, noble gas, etc.). Some technical indications about these considerations are given by Mazria (Mazria, 1975). Though two levels experimental design, (Zalewski, 1997) shows that external glazing emissivity, window type and wall absorptivity mainly characterize the system.

Trombe-Michel system (and its variants) inspired numerous publications about temperate climate and proved their efficiency as passive heating system. By extension, there is also a lot of studies about similar system, as solar chimney (Ong, 2003), (Awbi, 2003).

There are several values usually used to describe Trombe-Michel wall functioning:
- Transmitted energy through the wall by conduction,
- Transmitted energy through the holes by sensitive enthalpy,
- The system efficiency, defined as the rate between the energy received into the room over the total energy received by the North wall (South hemisphere) in the meantime.
- The FGS (solar benefits fraction), defined as the rate between the benefits from the Trombe-Michel system and the energy needed by the room without it, for a room temperature of 22°C.

*Nota:* Even if the FGS refers to active heating systems, it allows to compare easily passive heating systems between them and to other installations.

### 4.2.2 Presentation of Trombe-Michel wall simulations under *CODYRUN*

This numerical study is resolved by software simulations. The case study case presented here takes place in Antananarivo, Madagascar (Indian Ocean). The aim of the simulations conducted was to help this developing country to face low temperatures in classrooms during winter season.

The word 'zone', used previously, is understood as *CODYRUN* vocable, correspond usually to one room. Into the actual description, isothermal air assumption is made into the capitation area. This consideration is taken as first approximation and can be possibly modified later to improve the model (e.g. replaced by linear gradient hypothesis, etc.). According to the bibliography, one of the main issues mentioned is about the convection coefficient into the Trombe side of the wall. Because of the system successively laminar and turbulent, convection correlation such as - $Nu = f(Gr)$ – cannot be used. In this case, (Zalewski, 1997) says that the laminar coefficient evolve between 2 and 2.3 $W.m^{-2}.K^{-1}$ and the turbulent one evolve between 2.25 and 3.75 $W.m^{-2}.K^{-1}$. So, in a first approach, we approximate the temperature-evolving value of the convection coefficient as the average of the turbulent and laminar values (2.9 $W.m^{-2}.K^{-1}$).

Considering thermo-circulation, the *CODYRUN*'s airflow module allows to calculate air flows through wall's holes. This consideration and the case study description, the two areas (the Trombe system and the room) are studied coupled, which is a right physical representation of the reality. Results validity are certified by the fact that each module of *CODYRUN*, and their combinations, has been validated by comparison with experimental results, reference codes and BESTEST procedures.

This multi-model code structure (Boyer, 1996) can be efficiently exploited by studies such as this one, where it is necessary to choose the convection coefficients by area or by wall, to mesh slightly the walls to calculate precisely the conductive fluxes, to choose a solar gains repartition model or equivalent sky temperature, etc.

*CODYRUN* also creates numerous output simulation variables, in addition to the one related to the dry temperature area, such as sensible power outputs from mass exchange through wall holes, conductive fluxes through the wall, enthalpic zone balance and the PMV (Predictive Mean Vote, according to Gagge) of interior zone to quantify comfort conditions. It is also possible to explore the conductive flux through the Trombe-Michel wall glass.

Simulation results are presented with meteorological data of Antananarivo coming from TRNSYS 16.0 software and its TMY2 documentation data file. According to this one, daily sunshine insulation annual average is about 5.5 kWh.m$^{-2}$. The one on the North and East walls are respectively 2.23 and 2.67 kWh.m$^{-2}$. Modified in function of the current albedo (0.3), the previous values become 3 and 3.46 kWh.m$^{-2}$.

The simulated building represents a residence unit based on a geometric cell description (9 m² parallelepiped square base ground). Moreover, materials, thermo-physics characteristics and wall type are typical from Madagascar. The room is supposed to be constantly occupied by two persons and have a usual profile for intern lighting. This area is supposed to exchange air with the outside at a rate of 1 volume per hour. Finally, the system area is of 2 m² and is located on the North wall (South hemisphere).

### 4.2.3 Simulation

Passive heating oriented-simulations are made the first day of July (Julian calender days 181 and following). In a first time, the objective is to explain the set-up functioning. So, some hourly based analyses will be conducted. Finally, the last experiment will compare the same Trombe-Michel wall with and without the recycling variant.

#### 4.2.3.1 Inverse thermo-circulation evidence

By night, over some system specific conditions (open holes) and in function of previous day insulations (e.g. low insulation), the flow evolution can sometime be the symmetric of the one by day (Fig. 13). In this configuration, the system functioning is reversed, cooling the room instead of heating it. Between points 19 and 32 (i.e. 7 pm to 8 am next day), the Trombe-area temperature is lower than the room temperature, inducing the inverse thermo-circulation. This physical process can be illustrated by the negative part of the sensitive aeraulic benefits for the inside area.

Even if the flux values by night are low, it is important to note the following point : the curves are obtained by assuming a sky temperature equal to the outside air temperature. This assumption reduces the impact of the system/sky radiation transfer, so involves an error. This error could be reduced by taking a sky temperature value more realistic, so by considering the night-sky wide wavelength emission and the high glass emissivity (0.9). A better approximation can be done by a sky temperature model such as $T_{sky}=T_{ae}- K$, where $K=6$.

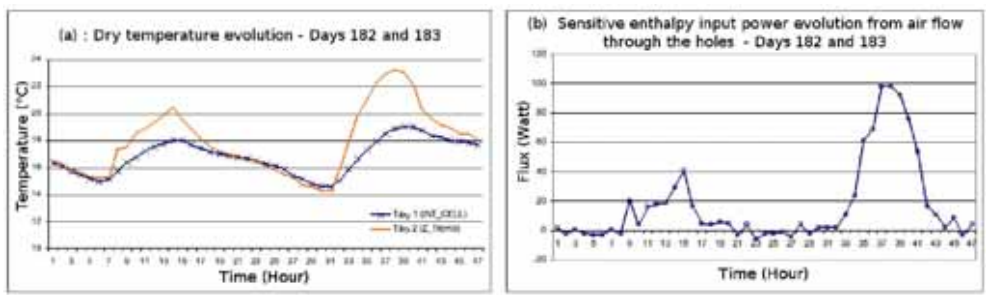

Fig. 13. Dry temperatures and sensible power balance through holes.

#### 4.2.3.2 Night hole closing system

Hole closing system strategy deserve a specific comment. This one can be done nicely from flow measurement, but a simpler and passive way to do it also exists. This one consists in a plastic foil, only attached on one side and occulting the holes preventing from inverse thermo-circulation. Into *CODYRUN*, the closure system is designed by the "little opening" component added to a hourly profile (i.e. 24 opening percentages), where the hole is closed by night and opened by day. However, the thermo-circulation (by system inertia) continues during the two first hours of the evening, so the profile includes this pattern too. Because of the small gap with the previous curves (Fig. 13), the results are firstly surprising. An explanation can be done by the small thermal gradient during the considered night. According to Mazria (Mazria, 1975) and Bansal (Bansal, 1994), recycling variant are only useful into cold climates (winter averages: -1 °C to -7 °C). So, simulation results show the minor impact of the hole control in this study case. In other words, it seems that the mass flow have a very low sensibility to the opening state. This hypothesis is rejoined by another author (Zalewski, 1997) who have substantively the same conclusion for a North France located case.

**4.2.3.3 Global Trombe wall functioning**

To obtain the full power delivered by the wall, airflow and conductive heat transfer have to be summed. By night, it appears a constantly positive heat flux inducing a rise-up of the room temperature (Fig. 14). In the morning (e.g. 9am first day), the global flux becomes negative. A full curve interpretation is approached by airflow and conductive heat flux analyze through the wall (right part of Fig. 14). Early morning, the global heat flux inversion can be explained by highly negative conductive heat flux resulting certainly from the heat flux night inversion (around 9 to 14, so a delay of 4 hours). Drawn curves of the incoming heat flux through the wall confirm this hypothesis.

As foreseeable (without simulation tools), the curves show the alternate functioning of the conductive and sensible airflow input power (the wall characteristics setting the shifting period). By night, the conductive heat flux through the wall rises-up the room temperature, whereas by day this action is done by the heat flux associated to air displacement, physically expressed by the natural air movement which is *a priori* synchronous with the sunlight distribution profile.

The previous report allows to ask about the holes necessity. In fact, even if there is no heat transfer improvement, the main interest comes from the heat distribution and its time-evolving modulation into the building.

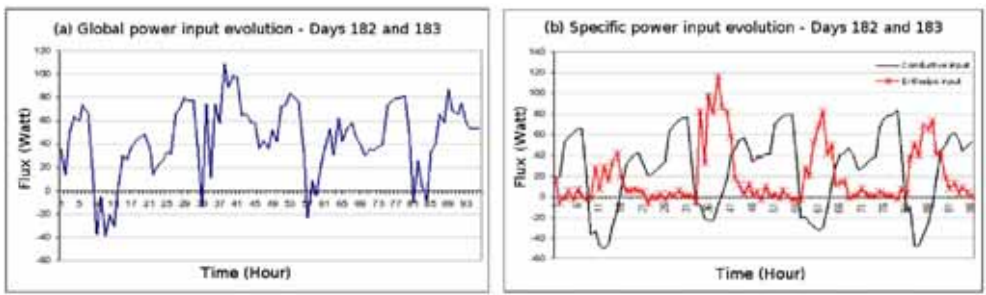

Fig. 14. Global and specific fluxes induced by the Trombe wall into the room.

To increase the room input power (i.e. the global heat flux), it could be simple to rise-up the wall surface and/or to add some reflectors to raise the incoming sunlight intensity.

However, in every cases the system have to be simulated for summer conditions. In order to reduce the external loss (illustrated by negative heat fluxes), a double glazing could be used, even if it reduces the incoming sunlight intensity (transmission loss). For reference, double glazing can be easily made from simple glazing. Another solution consists in isolating the indoor side of the wall.

**4.2.3.4 Trombe-Michel wall and recycling variant comparison**

The comparison, in the meantime, of the wall's holes presence (or not), expresses a gap between the global heat fluxes. The values corresponding to the recycling variant are shifted to the left compared to the classic Trombe-Michel wall (Fig. 15). The input power delivered with the holes is slightly superior, which confirms the hypothesis of low impact on section 4.2.3.2. A piece of explanation (checkable by simulation) can be given by the thermal gradient between the room and the Trombe system, the decrease of this one reducing the conductive transfer through the glazing.

Predicted Mean Votes (PMV) curves (of the right of Fig. 15) express the comfort sensation which evolve between -3 and +3. By comparison, it can be affirmed that the holes improve the comfort sensation. This improvement is related to the previous comment (about the superior input power amount which led to a rise-up of the dry temperature), and to the higher indoor side wall temperatures.

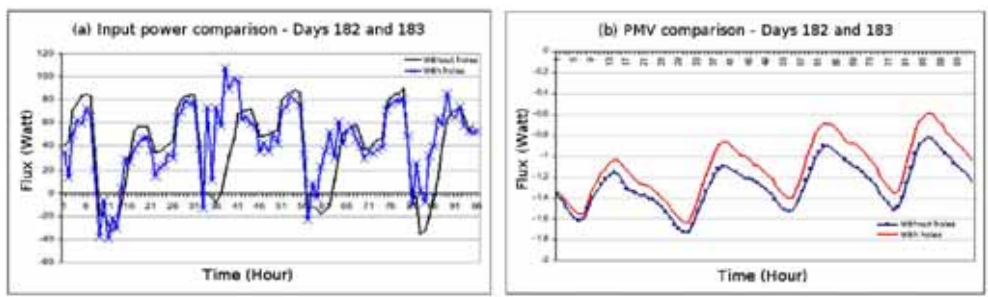

Fig. 15. Input power and PMV comparisons

**4.2.4 Trombe wall simulation exercise conclusion**

This example of *CODYRUN* use allows to find some interesting results about Trombe-Michel wall configurations for the specific location of Antananarivo. Results such as the interest of the recycling variant (which is almost not influenced by hole closing procedure), the time-repartition (and impact) of the airflow and conductive heat flow on the room temperature, (etc.) have been found. Considering the case study, much more informations could be deduced from other well designed simulations. The considerable amount of knowledge found was only deduced from simulations, so only following a numerical approach. From this consideration and the results precision, a good estimation about *CODYRUN* quality and level of development can be done.

## 5. Conclusion

This paper describes the models, validation elements and two applications of a general software simulation of building envelopes, *CODYRUN*. Many studies and applications have been conducted using this code and illustrate the benefits of this specific development.

Conducted and sustained for over fifteen years, the specific developments made around *CODYRUN* led our team to be owner of a building simulation environment and develop connex themes such as validation or application to large scale in building architectural requirements. This evolving environment has enabled us to drive developments in methodological areas (Mara, 2001) (modal reduction, sensitivity analysis, coupling with genetic algorithms (Lauret, 2005), neural networks, meta models,…) or related to technological aspects (solar masks, integration of split-system, taking into account the radiant barriers products, …). This approach allows a detailed understanding of the physical phenomena involved, the use of this application in teaching and capitalization within a work tool of a research team. Accompanied by the growth of computer capabilities, the subject of high environmental quality in buildings, our field becomes more complex by the integration of aspects other than those originally associated with the thermal aspects (environmental quality, including lighting, pollutants, acoustics, …). Simultaneously, it is also essential to achieve a better efficiency in the transfer of knowledge and operational tools.